\documentclass[10pt,twoside]{hsqcd}
\usepackage{epsf,amsmath}
\usepackage{graphicx}

\begin{document}

\title{LONGITUDINAL LOOP INTEGRALS IN THE GAUGE INVARIANT EFFECTIVE ACTION FOR HIGH ENERGY QCD
\thanks{This work is
supported by the Graduiertenkolleg "Zuk\"unftige Entwicklungen in der Teilchenphysik" }}

\author{ \underline{M. Hentschinski}, J. Bartels and L.N. Lipatov \\ \\
II. Institute for Theoretical Physics, Universit\"at Hamburg \\ 
22761 Hamburg, Germany \\
E-mail: martin.hentschinski@desy.de }

\maketitle

\begin{abstract}
\noindent We study integrations over light-cone degrees of freedom in the
gauge invariant effective action for high energy processes in QCD. We
propose a regularization which takes into account the signature of the
reggeized gluon. For a  test  we  apply it to the elastic
 and the production amplitude.
\end{abstract}



\markboth{\large \sl \underline{Hentschinski}, Bartels \& Lipatov 
\hspace*{2cm} HSQCD 2008} {\large \sl \hspace*{1cm} LONGITUDINAL INTEGRALS IN THE  EFFECTIVE ACTION }

\section{Introduction}
\label{sec:simple}

In 1995 an effective action \cite{action} for QCD scattering processes
at high center of mass energies $\sqrt{s}$ has been proposed by L.N.
Lipatov, which describes the interaction of fields of reggeized gluons
($A_\pm = -it^aA^a_\pm$ ) with quark ($\psi$) and gluon ($v_\mu =-it^a
v^a_\mu$) fields, local in rapidity. The effective action is given by
\begin{eqnarray}
  \label{eq:effact1}
  {S}_{\mbox{eff}} 
&=& \int d^4 x \big(
  \mathcal{L}_{\mbox{QCD}} (v_\mu, \psi) + 
\mbox{tr}[(A_-(v) -A_-)\partial^2 A_+ + (A_+(v) -A_+)\partial^2 A_-]\big)
 \nonumber \\
 \mbox{with}&&  A_\pm(v)  =v_\pm D_\pm^{-1}\partial_\pm = 
v_\pm - gv_\pm\frac{1}{\partial_\pm}v_\pm + g^2v_\pm\frac{1}{\partial_\pm}v_\pm\frac{1}{\partial_\pm}v_\pm - \ldots
\end{eqnarray}
In the above, light-cone components are defined by $k^\pm \equiv
n^\pm\cdot k$ where $n^{\pm}$ are the light cone directions associated
with the scattering particles. The fields of the reggeized gluon
obey the constraint $\partial_\pm A_\mp = 0$. The terms in
(\ref{eq:effact1}) which supplement the usual QCD-Lagrangian
$\mathcal{L}_{\mbox{QCD}}$ are called \emph{induced} terms and
consist, apart from the kinetic term of the reggeized gluon $-
\partial_\mu A^a_+ \partial_\mu A^a_-$ (the only part which is non-local in
rapidity), of the \emph{induced} vertices. For the production of real
particles within the Multi-Regge-Kinematics, it is then the induced
vertex $-gv_\pm\frac{1}{\partial_\pm}v_\pm \partial^2 A_\mp$ which
together with the three gluon vertex gives rise to the Lipatov
production vertex $C_\mu$.  Higher induced vertices are needed if the
production of real particles in the Quasi-Multi-Regge-Kinematics
(QMRK) is considered, as demonstrated in \cite{antonov}. From a point
of view of the QCD action, these induced vertices correspond to gluon
emissions from particles with a significantly different rapidity. As
(\ref{eq:effact1}) is local in rapidity, double-counting due to the
induced terms is absent for the real particle production amplitude.
\begin{figure}[tbp]
  \centering
  \begin{minipage}{.25\textwidth}
 \centering
\includegraphics[height=2cm]{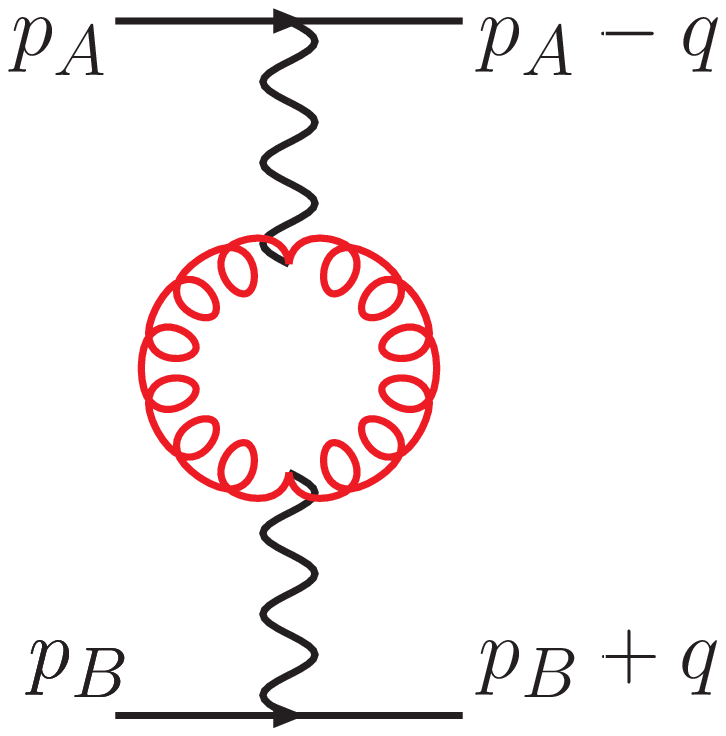} 
  \end{minipage}
\begin{minipage}{.1\textwidth}
  \end{minipage}
  \begin{minipage}{.25\textwidth}
 \centering
\includegraphics[height=2cm]{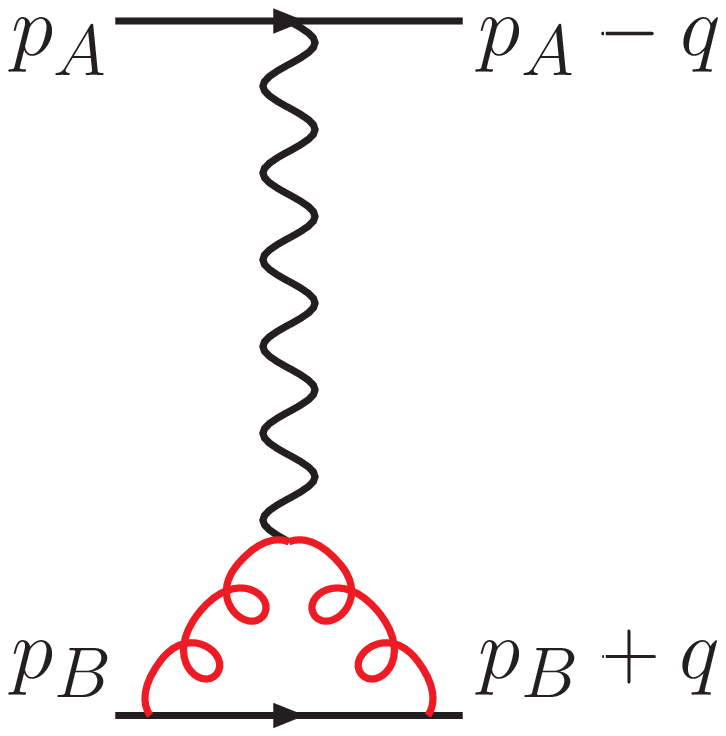} 
  \end{minipage}     
\begin{minipage}{.1\textwidth}
  \end{minipage}
\begin{minipage}{.25\textwidth}
 \centering
\includegraphics[height=2cm]{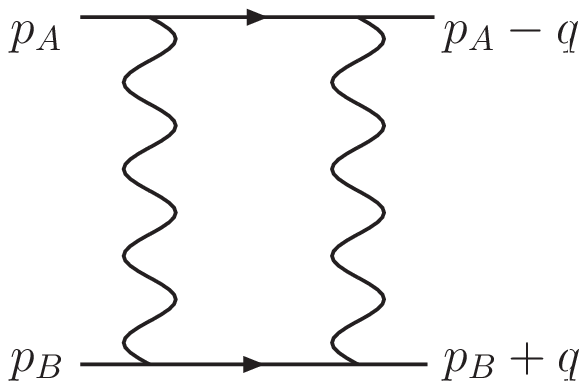} 
  \end{minipage}
 \\
\begin{minipage}{.25\textwidth}
  \caption{\small The central rapidity  correction. } 
  \label{fig:central}
  \end{minipage}
  \begin{minipage}{.1\textwidth}
  \end{minipage}
\begin{minipage}{.25\textwidth}
 \caption{\small Correction in the quasi-elastic region  } 
  \label{fig:quasi}
  \end{minipage}
\begin{minipage}{.1\textwidth}
  \end{minipage}
\begin{minipage}{.25\textwidth}
 \caption{\small Exchange of two reggeized gluons.  }
  \label{fig:two}
  \end{minipage}
\end{figure}
It seems appealing to use the effective action also to determine
virtual corrections to Reggeon-Reggeon and Reggeon-particle vertices.
Such corrections usually include an integration over light-cone
degrees of freedom. One finds that for the parts of the diagrams
connected with induced vertices, it is essential to impose some sort
of cut-off to ensure locality in rapidity of the integrand.  Moreover,
reggeized gluons are Reggeons which carry negative signature. 
Consequently, positive and negative signatured parts of an amplitude
are treated in different ways within the effective action.  We show in
the following how this can be done at the level of one-loop
integrations by considering the elastic (Sec.\ref{sec:elastic}) and
the production (Sec.\ref{sec:prodampl}) amplitude.

\section{The elastic amplitude}
\label{sec:elastic}
At one loop, the negative signatured part of the elastic amplitude is
solely given (including the imaginary part) by diagrams that represent
corrections to the exchange of one reggeized gluon i.e. quasi-elastic
(e.g. fig.\ref{fig:quasi}) and central rapidity (fig.\ref{fig:central})
corrections.  The latter leads to the dominant contribution
proportional to the logarithm in $s$ and will be discussed in the
following.

\subsection{The central-rapidity correction}
\label{sec:centralrap}

In the relevant contribution of fig.\ref{fig:central} the reggeized
gluon couples from above and from below by an induced vertex to the
gluon loop resulting into the following expression
\begin{equation}
  \label{eq:bare_centralrap}
i\mathcal{A}^{(-)}_{2 \to 2} (s, q_\perp^2) = s \Gamma^c_{AA'}  {2g^2 N_c}\int\!\! \frac{d^4 k}{(2\pi)^4} D_{{R}}(s_A, q_\perp^2)  \frac{{{q}}_\perp^2}{k^+} \frac{1}{k^2 (k-q)^2 } \frac{{{q}}_\perp^2}{k^-} D_{{R}}(s_B, q_\perp^2) \Gamma^c_{BB'},
\end{equation}
where $ \Gamma^c_{II'} = 2 ig t^c_{II'}$ is the coupling of the
reggeized gluon to the external quarks $I = A, B$ and $D_{{R}}(s_I,
q_\perp^2)$ the reggeized gluon's propagator with $s_I = (p_I + k)^2$,
$I = A, B$ the sub-center-of-mass-energy squared of the (quark, gluon)
$\to$ (quark, gluon) sub-amplitudes. The poles $1/k^+$ and $1/k^-$
arise due to the induced vertices.  In the corresponding QCD-diagram,
the two gluons inside the loop couple directly to the quarks $A$ and
$B$ and it is then the propagator of both quarks that reduces for
large sub-center-of-mass-energies squared $s_A = p_A^+ k^-$ and $s_B =
p_B^- k^+$ to the poles $1/k^-$ and $1/k^+$ respectively. In order to
justify the use of the expression (\ref{eq:bare_centralrap}) it is
therefore necessary that the values of $s_A$ and $s_B$ are larger than
the typical transverse scale of the amplitude.  Further, as the quarks
interact with the gluon-loop by reggeized gluons, the corresponding
(quark, gluon) $\to$ (quark, gluon) sub-amplitudes should have
negative signature in $s_A$ and $s_B$ respectively.  The natural place
to implement both requirements is the Reggeon propagator, as it is the
only part of the effective action which is non-local in rapidity.
Using the following definition, the negative signature constraint is
there easily implemented. We define
\begin{equation}
  \label{eq:prop}
  D_{{R}} (s_I, q^2_\perp) \equiv \frac{-i/2}{q^2_\perp} \lim_{\nu \to 0}\int_{-i\infty}^{i\infty} \frac{d \omega}{4\pi i} \frac{1}{\omega + \nu} \bigg[\bigg(\frac{-s_I}{\Lambda_I} \bigg)^{\omega_i} +\bigg(\frac{s_I}{\Lambda_I} \bigg)^{\omega_i} \bigg] = \theta(s_I - \Lambda_I)\frac{-i/2}{q^2_\perp}
\end{equation}
where $\Lambda_I$, $I = A, B$ is a scale significantly smaller than
$s$, but larger than the typical transverse scale of the elastic
amplitude. Note that $-\nu$ has the interpretation of an
infinitesimal small Regge-trajectory.  Inserting (\ref{eq:prop}) into
(\ref{eq:bare_centralrap}) and performing the integration over $k^+$
and $k^-$ we obtain
\begin{equation}
  \label{eq:elastic_negsig_result}
  i\mathcal{A}^{(-)}_{2 \to 2} (s,  q_\perp^2) =   s \Gamma^c_{AA'} \frac{1}{2}\bigg[\ln\big(\frac{-s{{k}}_\perp^2}{\Lambda_A\Lambda_B}\big)+ \ln\big(\frac{s{{k}}_\perp^2}{\Lambda_A\Lambda_B}\big) \bigg] \beta(q) \frac{-i/2}{q^2_\perp} \Gamma^c_{BB'} 
\end{equation}
which apparently has negative signature in $s$ as required.  $\beta(q)$
is the well-known gluon trajectory function. Note, that
(\ref{eq:elastic_negsig_result}) contains due to $\ln(-s) = \ln |s| -
i\pi$ also the corresponding imaginary part of the elastic amplitude.
The dependence on the scales $\Lambda_A$ and $\Lambda_B$ in
(\ref{eq:elastic_negsig_result}) can be shown to cancel with similar
contributions arising from the quasi-elastic regions (e.g.
fig.\ref{fig:quasi}) with rapidities close to the scattering particles
$A$ and $B$.

\subsection{Exchange of two reggeized gluons}
\label{sec:tworeggeons}

 For diagrams that contain two reggeized gluons in the $t$-channel (e.g. fig.\ref{fig:two}), the integrals over longitudinal loop
momenta $k^+$ and $k^-$ factorizes, and it is sufficient to consider
the 2 quarks-2 Reggeons vertex given by
  \begin{equation}
    \label{eq:tworegg_def}
\mathcal{A}_I^{a_1a_2} = {g^2 2\sqrt{2 }} \int \frac{d s_I}{2\pi i} \bigg(\frac{(t^{a_1}t^{a_2})_{II'}}{-s_I + k^2_\perp + i\epsilon} + \frac{(t^{a_2}t^{a_1})_{II'}}{s_I + (k-q)^2_\perp + i\epsilon} \bigg).
  \end{equation}
  The exchange of two reggeized gluons should yield now the
  positive signature part of the elastic amplitude. Indeed, for
  symmetric color quantum numbers which correspond to positive
  signature exchange, the integral over $s_I$, $I= A, B$ with $s_A =
  p_A^+k^-$ and $s_B = p_B^-k^+$ is convergent and we obtain the
  well-known QCD result.  For anti-symmetric color quantum numbers
  corresponding to negative signature exchange the integral in
  (\ref{eq:tworegg_def}) turns out to be divergent for large values of
  $s_I$.  However as demonstrated in Sec.\ref{sec:centralrap},
  contributions with large $s_I$ and negative signature are taken into
  account by the central and quasi-elastic contributions.  One
  therefore should isolate the part of the integral that was already
  contained in those contributions and cut off  the integral
  for values of $s_I$ being larger than the scale $\Lambda_{A,B}$
  introduced in the previous paragraph. This leads to convergence and
  furthermore vanishing of the two Reggeon exchange contribution with
  negative signature in accordance with the result that the negative
  signatured part of the elastic amplitude is described by the
  exchange of a single reggeized gluon alone.

\section{The production amplitude}
\label{sec:prodampl}
\begin{figure}[tbp]
  \centering
  \includegraphics[height=4cm]{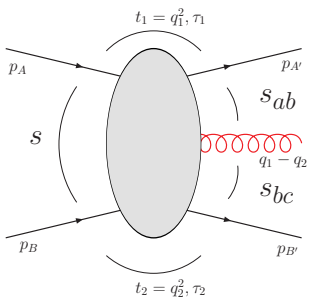} 
  \caption{The five point amplitude with $s = (p_A + p_B)^2$, $s_{ab}= (p_A' + k)^2$, $s_{bc} = (p_B' + k)^2$ and $k = q_1 - q_2$}
  \label{fig:fivepoint}
\end{figure}
The double-Regge limit of the five-point amplitude $\mathcal{A}_{2 \to3}$ (fig
\ref{fig:fivepoint}),  where $s, s_{ab}, s_{bc} \to \infty$,
$s_{ab}/s, s_{bc}/s \to 0$ and $\eta = s_{ab} s_{bc}/s, t_1, t_2$ are fixed,
 is known to possess the following analytic representation with $\tau_{i}$, $i=1,2$ the signature of the corresponding 
$t$-channel:  
\begin{eqnarray}
  \label{eq:analytic_struct}
{\cal A}_{2 \to 3}^{(\tau_1\tau_2)} = \int \frac{dj_1dj_2}{(2\pi i)^2} 
\big[
s^{j_1} s_{bc}^{j_2-j_1} \xi^{\tau_1}_{j_1}\xi^{\tau_2\tau_1}_{j_2j_1} F^L_{j_1j_2}( t_1, t_2, \eta) 
+
s^{j_2} s_{ab}^{j_1-j_2} \xi^{\tau_2}_{j_2}\xi^{\tau_1\tau_2}_{j_1j_2} F^R_{j_1j_2}( t_1, t_2, \eta) \big]
\nonumber \\
\mbox{with}  \qquad \xi^{\tau}_{j} = \frac{e^{-i\pi j} + \tau}{\sin \pi j}\qquad  \mbox{and} \qquad \xi^{\tau_1\tau_2}_{j_1j_2} =   \frac{e^{-i\pi(j_1-j_2)} + \tau_1\tau_2}{\sin \pi (j_1 -j_2)} \qquad \qquad \qquad \qquad
\end{eqnarray}
where $F^L_{j_1j_2}$ and $F^R_{j_1j_2}$ are real functions.  In
\cite{bartels} they have been determined to leading accuracy
considering discontinuities in $s$, $s_{ab}$ and $s_{bc}$. In the
effective action the different signature contributions to the
amplitude are determined as follows: For $\tau_1 = \tau_2 = +1$, one
simple inserts the Lipatov-vertex $C_\mu$ into the elastic amplitude
with positive signature exchange and one obtains the required
expression.  In the case where both t-channels carry different
signatures, one encounters the Reggeon-particle-2 Reggeons
vertex, which within the effective action was also addressed in
\cite{braun}.  Similar to Sec.\ref{sec:tworeggeons}, the integration
over $k^+$ and $k^-$ factorizes and can be carried out in a manner
similar to Sec.\ref{sec:tworeggeons}.  If both $t-$channels carry
negative signature, one has to determine the one-loop correction to
the Lipatov production vertex $C_\mu$, which in principle involves
next-to-leading accuracy calculations.  Here we evaluate the resulting
integrals using the methods of Sec.\ref{sec:centralrap} but restrict
ourselves to the leading logarithmic part, while we keep track of all
imaginary parts. For every combination of signatures, our result turns
out to be in accordance with the one of \cite{bartels}.

\section{Concluding remarks}
\label{sec:concl}
We gave a prescription how longitudinal integrals at one loop can
consistently be performed in the effective action. The prescription
has been tested with the elastic and the production amplitude and we could
reproduce correctly the leading logarithms and the
corresponding energy discontinuities.

\end{document}